\documentclass[10pt, letter]{bmc_article}    

\usepackage{cite} 
\usepackage{url}  
\usepackage{ifthen}  
\usepackage{multicol}   
\usepackage[utf8]{inputenc} 
\usepackage{amsmath, amssymb}
\usepackage{graphicx,bigstrut,color}
\usepackage{booktabs}
\usepackage{boxedminipage}
\usepackage{nameref}

\newboolean{publ}

\begin{document}

\title{Operation of a Brain-Computer Interface Walking Simulator by Users with Spinal Cord Injury}
 
\author{Christine E. King\correspondingauthor$^{1}$%
         \email{Christine E. King\correspondingauthor - kingce@uci.edu}%
         \and 
         Po T. Wang$^{1}$%
         \email{Po T. Wang - ptwang@uci.edu}%
         \and
         Luis A. Chui$^{2}$%
         \email{Luis A. Chui - lchui@uci.edu}%
         \and
         An H. Do\correspondingauthor$^{2, 3}$%
         \email{An H. Do\correspondingauthor - and@uci.edu}%
         \and
         Zoran Nenadic$^{1, 4}$%
         \email{Zoran Nenadic - znenadic@uci.edu}
      }

\address{%
    \iid(1)Department of Biomedical Engineering, University of California, Irvine, CA 92697 USA\\
    \iid(2)Department of Neurology, University of California, Irvine, CA 92697 USA\\
		\iid(3)Division of Neurology, Long Beach Veterans Affairs Medical Center, Long Beach, CA 90822 USA\\
                \iid(4)Department of Electrical Engineering and Computer Science, University of California, Irvine, CA 92697 USA
}%

\maketitle

\begin{abstract}
\paragraph*{Background:} Spinal cord injury (SCI) can leave the affected individuals with paraparesis or paraplegia, thus rendering them unable to ambulate. Since there are currently no restorative treatments for this population, novel approaches such as brain-controlled prostheses have been sought. Our recent studies show that a brain-computer interface (BCI) can be used to control ambulation within a virtual reality environment (VRE), suggesting that a BCI-controlled lower extremity prosthesis for ambulation may be feasible. However, the operability of our BCI has not yet been tested in a SCI population.
   
\paragraph*{Methods:} Five subjects with paraplegia or tetraplegia due to SCI underwent a 10-min training session in which they alternated between kinesthetic motor imagery (KMI) of idling and walking while their electroencephalogram (EEG) were recorded. Subjects then performed a goal-oriented online task, where they utilized KMI to control the linear ambulation of an avatar while making 10 sequential stops at designated points within the VRE. Multiple online trials were performed in a single day, and this procedure was repeated across 5 experimental days.      
  
\paragraph*{Results:} Classification accuracy of idling and walking was estimated offline and ranged from 60.5\% (p=0.0176) to 92.3\% (p=1.36$\times$10$^{\text{-20}}$) across subjects and days. Offline analysis revealed that the activation of mid-frontal areas mostly in the $\mu$ and low $\beta$ bands was the most consistent feature for differentiating between idling and walking KMI. In the online task, subjects achieved an average performance of 7.4$\pm$2.3 successful stops in 273$\pm$51 sec. These performances were purposeful, i.e. significantly different from the random walk Monte Carlo simulations (p$<$0.01), and all but one subject achieved purposeful control within the first day of the experiments. Finally, all subjects were able to maintain purposeful control throughout the study, and their online performances improved over time.  

\paragraph*{Conclusions:} The results of this study demonstrate that SCI subjects can purposefully operate a self-paced BCI walking simulator to complete a goal-oriented ambulation task. The operation of the proposed BCI system requires short training, is intuitive, and robust against subject-to-subject and day-to-day neurophysiological variations. These findings indicate that BCI-controlled lower extremity prostheses for gait rehabilitation or restoration after SCI may be feasible in the future.

\end{abstract}

\section*{Keywords}
Spinal cord injury, brain computer interface, virtual reality environment, electroencephalogram, kinesthetic motor imagery, gait, ambulation, locomotion.

\ifthenelse{\boolean{publ}}{\begin{multicols}{2}}{}

\newpage
\section*{Introduction}
Spinal cord injury (SCI) can leave the affected individuals with paraparesis or paraplegia, thus rendering them unable to ambulate. Since there are currently no restorative treatments for this population, technological approaches have been sought to substitute for the lost motor functions. Examples include robotic exoskeletons~\cite{argo:10}, functional electrical stimulation (FES) systems~\cite{dgraupe:98}, and spinal cord stimulators~\cite{sharkema:11}. However, these systems lack the able-body-like supraspinal control, and so the ambulation function of these devices is controlled manually. In addition to being unintuitive, these systems may be costly and cumbersome to use, and therefore have not yet garnered popular appeal and adoption among potential SCI users. Due to these limitations, wheelchairs remain the primary means of mobility after SCI. Unfortunately, the extended reliance on wheelchairs typically lead to a wide variety of comorbidities that constitute the bulk of chronic SCI-related medical care costs~\cite{jkschmitt:03, ssabharwal:03, yzehnder:04, mmpriebe:03}. Consequently, to address the above issues associated with the treatment of paraparesis and paraplegia after SCI, novel brain-controlled prostheses are currently being pursued~\cite{ptwang:10, ptwang:12}. 

Recent results by our group~\cite{ptwang:10, ptwang:12} suggest that an electroencephalogram (EEG) based brain-computer interface (BCI) controlled lower extremity prosthesis may be feasible. More specifically, these studies demonstrated the successful implementation of a BCI system that controls the ambulation of an avatar (a stand-in for a lower extremity prosthesis) within a virtual reality environment (VRE). By using a data-driven machine learning approach to decode the users' kinesthetic motor imageries (KMIs), this BCI-controlled walking simulator enabled a small group of subjects (one with paraplegia due to SCI~\cite{ptwang:12}) to achieve intuitive and purposeful BCI control after a short training session. While the single SCI subject outperformed most able-bodied subjects in this study, the operability of this system has not yet been tested in a SCI population. The successful implementation of the BCI-controlled walking simulator in a population of subjects with SCI will establish the feasibility of future BCI-controlled lower extremity prostheses and will represent an important step toward developing novel gait rehabilitation strategies for SCI.  

Extending the application of the BCI-controlled walking simulator to a SCI population is faced with several problems. First, cortical reorganization, which is common after SCI~\cite{Cramer2005, Sabbah2002, Alkadhi2005, Hotz-Boendermaker2008}, may cause the cortical representation of walking KMI to vary vastly from one SCI subject to another. Second, this representation may dramatically evolve over time when SCI subjects are engaged in KMI training~\cite{Cramer2007}. Finally, subjects with SCI may interpret walking KMI either as motor imagery or as attempted walking, which in turn may result in multiple patterns of cortical activation across these individuals. Therefore, intuitive BCI operation under these conditions requires a system that can accommodate for the variations of brain physiology across SCI individuals, time, and strategies. To address these problems, we used a data-driven machine learning method to decode walking KMIs in a small population of SCI individuals. This approach enabled 5 subjects to achieve intuitive and self-paced operation of the BCI-controlled walking simulator after only minimal training. Furthermore, they were able to maintain this level of control over the course of several weeks. 

\section*{Methods}
\subsection*{Overview}
The goal of this study was to determine if individuals with complete motor SCI can use intuitive control strategies to purposefully operate a BCI-controlled walking simulator. To achieve this goal, 5 subjects with SCI underwent a short training procedure where they performed alternating epochs of idling and walking KMI while their EEG were recorded. These training EEG data were then analyzed to build decoding models for subsequent online BCI operation. To ascertain purposeful BCI control, subjects then performed 5 sessions of an online BCI goal-oriented virtual walking task~\cite{ptwang:12}. This entire procedure was performed 5 times over the course of several weeks to determine if subjects' performances improved with additional practice. 

\subsection*{Subject Recruitment}
This study was approved by the University of California, Irvine Institutional Review Board. Four subjects with paraplegia and one with tetraplegia due to SCI were recruited via physician referral from the Long Beach Veterans Affairs Spinal Cord Injury Center and other SCI outreach programs. The subjects (see Table 1) gave their informed consent to participate in the study. Note that all subjects were BCI na\"{i}ve and most of them performed the experimental procedures at a rate of once per week.

\begin{table}[htbp]
	\centering
			\begin{tabular}{cccl}
\hline \hline
Subject & Gender & Age & SCI status \\
\hline
1 & M & 34  & T11, ASIA A, 8 yr. post injury  \\
2 & M & 46  & T1, ASIA B, 4 yr. post injury \\
3 & M & 43  & C5, Syringomyelia, 14 yr. post onset \\
4 & M & 59  & T1, ASIA B, 2 yr. post injury \\
5 & M & 21  & T11, ASIA B, 1 yr. post injury \\
\hline \hline
\end{tabular}
	\caption{List of participants with demographic data and level of SCI. ASIA = American Spinal Injury Association impairment scale.}
	\label{tab:table1}
\end{table}

\subsection*{Data Acquisition} 
Each subject was positioned $\sim$0.8 to 1 m from a computer screen that displayed textual cues or the VRE. EEG were recorded using a 63-channel cap (Medi Factory, Heerlen, The Netherlands) with Ag-AgCl electrodes arranged according to the extended 10-20 International Standard.  Conductive gel (Compumedics USA, Charlotte, NC) was applied to all electrodes and the 30-Hz impedances between each electrode and the reference electrode were maintained at $<$10~K$\Omega$ by abrading the scalp with a blunt needle.  The EEG signals were amplified and digitized (sampling rate: 256 Hz, resolution: 22 bits) using two linked NeXus-32 EEG bioamplifiers (MindMedia, Roermond-Herten, The Netherlands). These signals were then streamed in real-time to a computer, re-referenced in a common average mode, band-pass filtered (0.01-40 Hz), and subsequently analyzed. The above procedures were performed using custom-written MATLAB (MathWorks, Natick, MA) programs. 

\subsection*{Training Procedure}\label{sub:training}
To facilitate intuitive control of ambulation within in a VRE, a EEG decoding model was developed that differentiates between idling and walking KMI. To this end, training EEG data were acquired while subjects underwent 30-sec alternating epochs of idling and walking KMI over a 10-min session. Subjects were instructed to perform walking KMI and idling via automated textual cues displayed on the screen. For the walking KMI task, subjects were coached to vividly imagine themselves walking or to attempt to perform the cyclic (albeit ineffective) leg movements of walking. During this entire procedure, a computer labeling signal (to identify idling and walking KMI epochs) was recorded by an auxiliary data acquisition system (MP150, Biopac Systems, Goleta, CA), and a common pulse train sent to both EEG and auxiliary data systems was used to synchronize the EEG and labeling signals. Also, the subjects were asked to remain still during the entire procedure, and their movement was monitored by the experimenter. If consistent movements were observed, the subject was asked to repeat the entire training procedure without making the undesired movements. 

\subsection*{Decoding Model Generation}\label{sub:osaapmg}

Offline signal analysis to generate EEG decoding models was performed in a manner similar to Wang et al.~\cite{ptwang:12}. Briefly, EEG and labeling signals were first aligned and merged using the synchronization signal, and an iterative artifact rejection algorithm~\cite{ahdo:11} was used to exclude EEG channels with excessive electromyogram (EMG) artifacts from further analysis. The pre-processed EEG data were then split into segments of idling and walking KMI states using the labeling signal. Each EEG segment was then divided into 5 randomly spaced, non-overlapping 4-sec trials, which were then transformed into the frequency domain using the Fast Fourier Transform (FFT). Note that the remaining data in each segment were discarded. Finally, the power spectral density of each trial was integrated in 2-Hz bins, resulting in 20 binned power spectral values per channel. 

Once the EEG segments were transformed into the frequency domain, a systematic search was performed to find the best contiguous frequency range for classification~\cite{ptwang:12}.  
The dimension of the input data was then reduced using classwise principal component analysis (CPCA)~\cite{Das2007, kdas:09}, and discriminating features were extracted using either Fisher's linear discriminant analysis (LDA)~\cite{roduda:01} or approximate information discriminant analysis (AIDA)~\cite{znenadic:07, Das2008}. This resulted in the extraction of one-dimensional (1D) spatio-spectral features:

\begin{equation}\label{eq:features}
f = \mathbf{T}\mathbf{\Phi}_{\text{C}}(\mathbf{d})
\end{equation}  
where $\mathbf{d}\in \mathbb{R}^{B \times C}$ is a single-trial of EEG data with $B$ frequency bins per channel and $C$ retained EEG channels, $\mathbf{\Phi}_{\text{C}}:\mathbb{R}^{B \times C}\rightarrow \mathbb{R}^{m}$ is a piecewise linear mapping from the data space to an $m$-dimensional CPCA-subspace, and $\mathbf{T}:\mathbb{R}^{m}\rightarrow \mathbb{R}$ is a LDA or AIDA transformation matrix. 

Once the single-trial data were reduced to 1D spatio-spectral features, a linear Bayesian classifier: 
\begin{equation}
\label{eq:lrt}
f^{\star} \in
\begin{cases}
\mathcal{I}, &\text{if}\quad P(\mathcal{I}\,|f^{\star})>P(\mathcal{W}\,|f^{\star})\\
\mathcal{W}, &\text{otherwise}
\end{cases}
\end{equation}
was designed in the feature domain, where $P(\mathcal{I}\,|f^{\star})$ and $P(\mathcal{W}\,|f^{\star})$ are the posterior probabilities of idling and walking classes given the observed feature, $f^{\star}$, respectively. Eq.~(\ref{eq:lrt}) can be read as: ``classify $f^{\star}$ as idle class if the posterior probability of idling is greater than the posterior probability of walking given the feature, $f^{\star}$,  and vice versa.'' The offline performance of the Bayesian classifier~(\ref{eq:lrt}), expressed as a classification accuracy, was then assessed by performing 10 runs of stratified 10-fold cross-validation (CV)~\cite{rkohavi:95}. To determine the optimal parameters for classification, the lower bound of the frequency range was then increased in 2-Hz steps, and this procedure was repeated until the classifier performance stopped improving, allowing the optimal lower frequency bound to be determined. The upper bound of the optimal frequency range was determined in a similar manner.
The classification accuracy was also used to decide whether to use AIDA or LDA as a feature extraction technique. Finally, the parameters of the decoding model, including the optimal frequency range, the feature extraction mapping, and the classifier parameters, were saved for real-time EEG analysis. 

\subsection*{Online Signal Analysis}\label{sub:osa}

During online operation, 0.5-sec long segments of EEG data were acquired in real time (at a rate of two segments per second), and were processed using the methods described in the \nameref{sub:osaapmg} section. Briefly, the EEG signals were first band-pass filtered, followed by automated artifact rejection. The pre-processed EEG data were then transformed into the frequency domain using FFT, and the power spectral densities over the optimal frequency range were calculated. These spectral data were used as an input for the feature extraction algorithm~(\ref{eq:features}). Finally, using Bayes' rule~(\ref{eq:lrt}), the posterior probabilities of idling and walking classes given the observed EEG feature, $f^{\star}$, were calculated. 

\subsection*{Online Calibration}\label{sec:oc}
Since the BCI system is a binary state machine with an idling and a walking state, a brief calibration procedure was required to identify the posterior probability thresholds to transition between these two states~\cite{ptwang:12}. Using the decoding model developed during the training procedure, the BCI system was set to run in the online mode while subjects were prompted to alternate between performing idling and walking KMI for $\sim$2 min. The posterior probabilities were recorded and their histograms were plotted to determine two thresholds, one to initiate the ambulation of the avatar, $T_{\mathcal{W}}$, and the other to stop the avatar, $T_{\mathcal{I}}$. These two thresholds were initially set as $T_{\mathcal{W}} = \text{median}\left\{P(\mathcal{W}\,|f^{\star}\in \mathcal{W})\right\}$ and $T_{\mathcal{I}} = \text{median}\left\{P(\mathcal{W}\,|f^{\star}\in \mathcal{I})\right\}$. Then, a brief online test was performed and based on the subject's feedback, the thresholds were adjusted as necessary. During online BCI operation, the posterior probabilities $P(\mathcal{W}\,|f^{\star})$ corresponding to the most recent 1.5 sec of EEG data were averaged and compared to the thresholds $T_{\mathcal{W}}$ and $T_{\mathcal{I}}$, and the state transitions were executed based on the finite state machine diagram given in Fig.~1. This step was implemented in order to smooth the sequence of posterior probabilities and reduce noisy state transitions.

\begin{figure}[htbp]
	\centering
		\includegraphics[width=\textwidth]{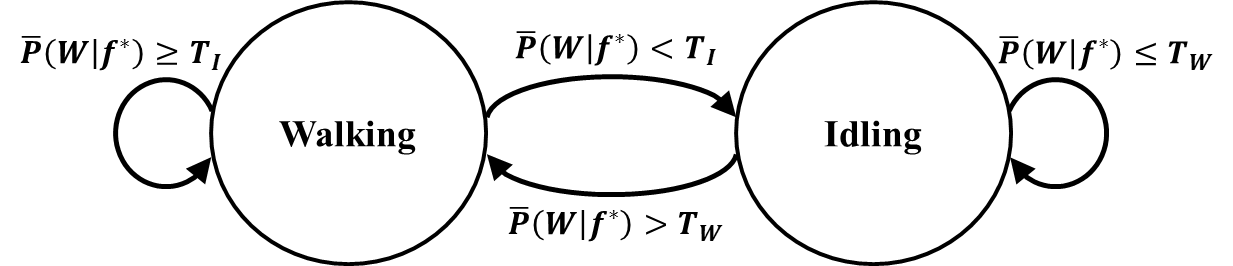}
	\caption{The BCI system as a binary state machine with walking and idling states represented by circles. The state transitions are represented by arrows, with transitions triggered by the conditions shown next to the arrows. Self-pointing arrows denote that the system remains in the present state.}
	\label{fig:Figure_1}
\end{figure}

\subsection*{BCI and Virtual Reality Environment Integration}
Using a simulated physics environment (Half-Life 2 Garry's Mod, Valve Corporation, Bellevue, WA), a VRE was constructed to represent a flat grassland with 10 non-player characters (NPCs) arranged in a linear path in front of an avatar (see Fig.~2). During online evaluation, the subjects were assigned the task of utilizing idling and walking KMI to control the avatar's standing and walking, similar to~\cite{ptwang:10,ptwang:12}. A virtual joystick program~\cite{DvanderWesthuysen:2011} was used to interface the BCI software (MATLAB) and the VRE to enable the BCI's control of the avatar's walking. Full details on this integration process can be found in Wang et al.~\cite{ptwang:12}.

\begin{figure}[htbp]
	\centering
		\includegraphics[width=\textwidth]{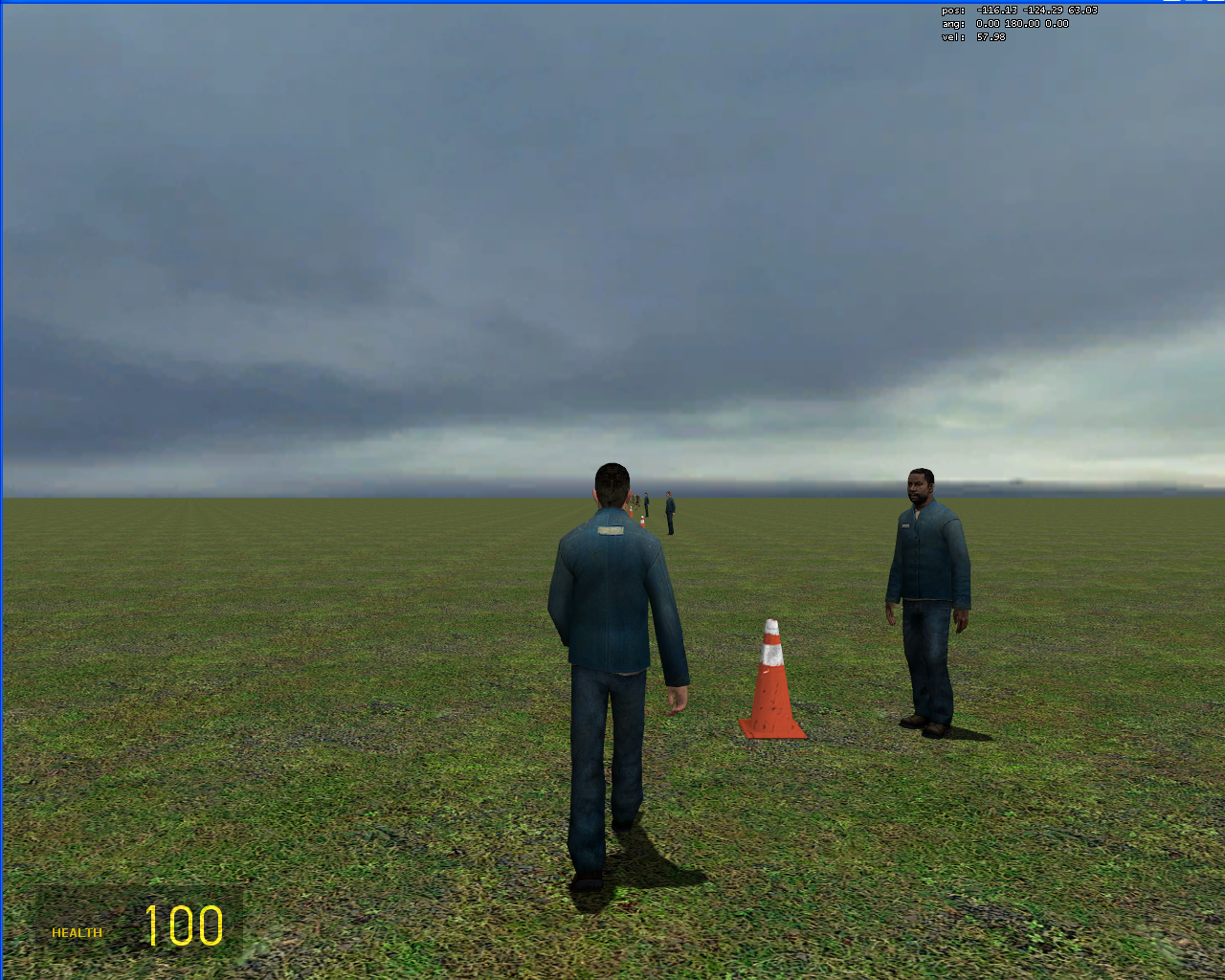}
	\caption{The avatar controlled by the BCI within the VRE. The simulator is operated in 3$^{\text{rd}}$ person view. The subject uses walking KMI to move the avatar to each NPC and idling KMI to dwell there for at least 2 sec.}
	\label{fig:Figure_2}
\end{figure}

\subsection*{Online Performance and Assessment}\label{sec:opaa}
To assess the online BCI performance, subjects were tasked to move the avatar within two body lengths of each NPC and remain idle at each location for at least 2 sec. On each experimental day, subjects repeated this task for 2--8 online sessions. This depended upon the subject's willingness and availability. In total, each subject underwent between 19 and 29 online sessions performed over 5 experimental days conducted over the course of several weeks. Two performance measures were recorded for each session: the number of successful stops at the NPCs and the time taken to complete the course. The successful stops were defined as follows. Subjects were given one point for dwelling at the designated stop for at least 2 sec, and a fraction of a point for dwelling at the designated stop for at least 0.5 sec. No point penalty was incurred for dwelling longer than 2 sec, but this inherently increased the time to completion. Thus, the maximum successful stop score was 10 points. In addition, subjects were given a 20-min time limit to complete the course. If the subject could not finish the course within 20 min, the trial was interrupted and the number of successful stops achieved thus far was recorded. 

\subsubsection*{Control Experiments}\label{sec:random}
The online performances (number of successful stops and completion time) were compared to random walk to determine if purposeful control was attained. The random walk performance was simulated by uniformly sampling the posterior probabilities between 0 and 1 and following the state transition rules with threshold values presented in \nameref{sec:oc} of the Methods section (details can be found in~\cite{ptwang:12}). To ensure statistical significance, 1000 Monte Carlo (MC) simulation runs were performed and the number of successful stops and completion times were logged (the 20-min completion time limit was enforced as above). In addition, the same task was performed by an able-bodied subject using a manually controlled physical joystick~\cite{ptwang:12}. 

\subsubsection*{Statistical Tests}\label{sec:statistics}
Purposeful control was defined as the ability to complete an online session within the allotted 20 min, with performances significantly different from random walk (p$<$0.01). This was ascertained by comparing the subjects' online performances to MC simulations using a multivariate analysis. To this end, the 2D probability density function (PDF) of each subject's simulated random walk performance (consisting of two variables: number of successful stops and completion time) was estimated using the Parzen-Rosenblatt window method~\cite{roduda:01, Botev2006}. A constant-value PDF contour was then drawn through each subject's observed performance point and the corresponding p-value was determined as the fraction of the volume under the random walk PDF outside of this contour. These p-values were used to ascertain a statistically significant difference between the online performances and the simulated random walk. Finally, subjects are deemed to have attained purposeful control once they have a single online session that is purposeful based on the definition above.

\subsubsection*{Composite Score}\label{sec:cs}
Given the difficulty of interpreting multivariate performance measures across subjects and sessions, a single composite score was defined as the following geometric mean: 
\begin{equation}
\label{eq:composite2}
\begin{split}
c   &= \sqrt{c_s c_t}
\end{split}
\end{equation}
where $c_s$ and $c_t$ are the normalized performance measures for the number of successful stops and completion times, respectively, i.e.
\begin{equation}
\label{eq:composite1}
\begin{split}
c_s &= \frac{s}{s_{\text{max}}}   \\
c_t &= \frac{t_{\text{max}}-t}{t_{\text{max}}-t_{\text{min}}} \\
\end{split}
\end{equation}
Here, $s$ is the subject's number of successful stops,  $s_{\text{max}} = 10$ is the maximum number of successful stops,  $t$ is the subject's completion time, $t_{\text{max}} = 1200$ sec is the maximum allowed time,  and $t_{\text{min}} = 201.52$ sec is the minimum time required to complete the course while achieving 10 successful stops. The values of $c_s$ and $c_t$, and consequently $c$, range from 0 to 100\%, where 100\% corresponds to a perfect performance. Note that the use of the geometric mean favors a performance that is balanced over a performance that sacrifices one performance measure over the other (e.g. when a subject finishes the course in a short time while failing to make stops). Also note that the normalization of $c_s$ and $c_t$ ensures that the performance measures are unitless.

\section*{Results}

\subsection*{Offline Performances}\label{sec:offlineresults}
After subjects underwent the training data collection for each experimental day (see \nameref{sub:training} section),  subject-specific EEG decoding models were generated as described in the~\nameref{sub:osaapmg} section. Cross-validation of these models resulted in offline classification accuracies that ranged from 60.5\% to 92.3\% (Table 2),  with corresponding p-values between 0.018 and $10^{-20}$, thus indicating that the classifier performances were well above the chance level (50\%).  In addition, the subjects' average offline performances over 5 experimental days were 82.3\%, 71.8\%, 82.3\%, 82.5\%, and 82.2\%, respectively. Finally, the overall average for all 5 subjects across all 5 experimental days was 80.2\% $\pm$ 8.62 ($n=25$).

\begin{table}[htbp]
	\centering
\begin{tabular}{ccllcccll}
\toprule
 \multicolumn{4}{c}{Offline Performances} \\
\midrule
Subject & Day & $P(\text{correct}\,|f^{\star})$ (\%) & p-value \\
\midrule   
 1 & 1    & 71.9 $\pm$ 2.2 & $6.29\times10^{-6}$  \\ 
   & 2    & 89.4 $\pm$ 1.2 & $1.53\times10^{-17}$ \\
   & 3    & 83.9 $\pm$ 2.0 & $1.30\times10^{-12}$ \\
   & 4    & 84.0 $\pm$ 1.9 & $1.30\times10^{-12}$ \\
   & 5    & 82.2 $\pm$ 1.7 & $6.55\times10^{-12}$ \\
\cmidrule{2-4}
   & Avg. & 82.3 $\pm$ 6.4 & $1.26\times10^{-6}$  \\
\midrule   
 2 & 1    & 62.2 $\pm$ 1.8 & $6.00\times10^{-3}$  \\
   & 2    & 62.0 $\pm$ 1.8 & $1.05\times10^{-2}$  \\
   & 3    & 60.5 $\pm$ 2.0 & $1.76\times10^{-2}$  \\
   & 4    & 91.6 $\pm$ 1.7 & $1.60\times10^{-19}$ \\
   & 5    & 82.5 $\pm$ 1.6 & $6.55\times10^{-12}$ \\
\cmidrule{2-4}
   & Avg. & 71.8 $\pm$ 14.3 & $6.82\times10^{-3}$ \\
\midrule   
 3 & 1    & 90.3 $\pm$ 1.3 & $1.66\times10^{-18}$ \\
   & 2    & 83.9 $\pm$ 1.1 & $1.30\times10^{-12}$ \\
   & 3    & 72.8 $\pm$ 2.9 & $2.35\times10^{-6}$  \\
   & 4    & 81.0 $\pm$ 2.1 & $1.35\times10^{-10}$ \\
   & 5    & 83.3 $\pm$ 1.2 & $1.30\times10^{-12}$ \\
\cmidrule{2-4}
   & Avg. & 82.3 $\pm$ 6.3 & $4.69\times10^{-7}$  \\ 
\midrule
 4 & 1    & 74.7 $\pm$ 1.7 & $2.82\times10^{-7}$  \\
   & 2    & 92.3 $\pm$ 1.6 & $1.36\times10^{-20}$ \\
   & 3    & 81.5 $\pm$ 1.3 & $3.07\times10^{-11}$ \\
   & 4    & 80.5 $\pm$ 2.0 & $1.35\times10^{-10}$ \\
   & 5    & 83.5 $\pm$ 2.2 & $1.30\times10^{-12}$ \\
\cmidrule{2-4}
   & Avg. & 82.5 $\pm$ 6.4 & $5.64\times10^{-8}$  \\
\midrule
 5 & 1    & 82.7 $\pm$ 1.3 & $6.55\times10^{-12}$ \\
   & 2    & 86.3 $\pm$ 1.3 & $6.56\times10^{-15}$ \\
   & 3    & 78.9 $\pm$ 1.4 & $2.17\times10^{-9}$  \\
   & 4    & 82.2 $\pm$ 1.5 & $6.55\times10^{-12}$ \\
   & 5    & 81.0 $\pm$ 1.3 & $1.35\times10^{-10}$ \\
\cmidrule{2-4}
   & Avg. & 82.2 $\pm$ 2.7 & $4.30\times10^{-10}$ \\
\bottomrule 
\end{tabular}
	\caption{Offline performances represented as classification accuracies estimated using 10 runs of stratified 10-fold CV. The classification accuracy is defined as the probability of correctly classifying a trial given the feature, $f^{\star}$, i.e. $P(\text{correct}\,|f^{\star})= P(\mathcal{I}\,|f^{\star}\in\mathcal{I})P(\mathcal{I})+ P(\mathcal{W}\,|f^{\star}\in\mathcal{W})P(\mathcal{W})$, where $P(\mathcal{I}\,|f^{\star}\in\mathcal{I})$ and $P(\mathcal{W}\,|f^{\star}\in\mathcal{W})$ are defined in \nameref{sec:oc} in the Methods section, and $P(\mathcal{I})$ and $P(\mathcal{W})$ are the prior probabilities of idling and walking class, respectively.}
	\label{tab:table2}
\end{table}

Spatio-spectral feature extraction maps generated from offline analysis, as described in the~\nameref{sub:osaapmg} section, revealed salient brain areas and frequency bands underlying idling and walking KMI for each subject. Qualitative analysis of these topographic maps revealed significant variations in the brain areas utilized by the subjects while performing this task. For example, Subject 2 used mostly the Cz area (Fig.~3), whereas Subject 5 used areas C3 and C4 (Fig.~4). Furthermore, the utilized brain areas and frequency bands evolved for each subject over the 5 experimental days. For example, Subject 2 had growing activation of mid-frontal areas up to experimental day 4, followed by a slight shrinkage on day 5. However, one consistent feature among all subjects was the activation of mid-frontal areas (over Cz or FCz and adjacent electrodes) mostly in the $\mu$ (8--12 Hz) and low $\beta$ (12--16 Hz) bands.

\begin{figure}[htbp]
	\centering
		\includegraphics[width=\textwidth]{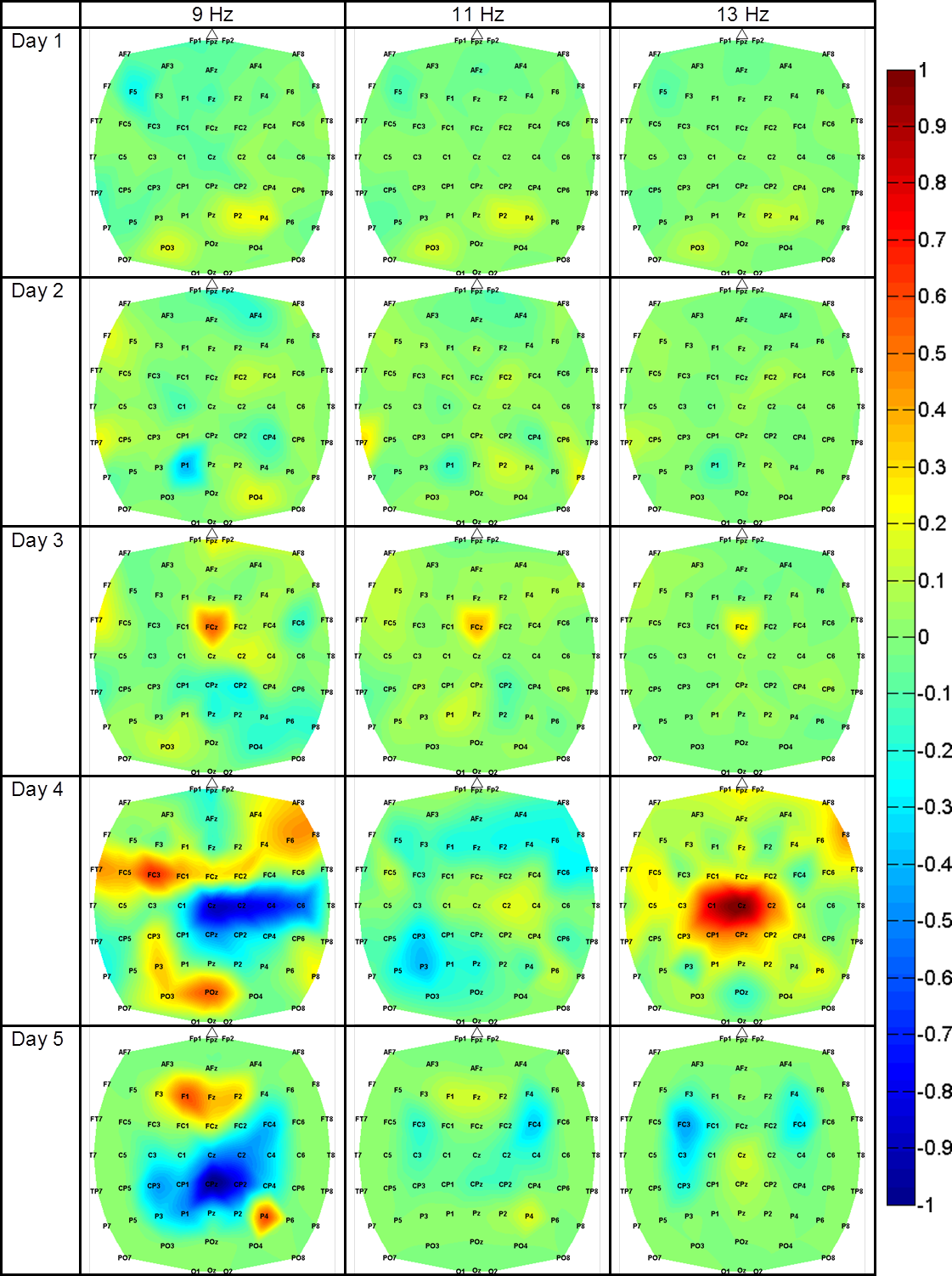}
	\caption{Feature extraction images of Subject 2 for all experimental sessions. Dark colors (red and blue) represent the areas that were responsible for encoding the differences between idling and walking KMI. The EEG power in the 9-13 Hz bins in the mid-frontal (FCz), central (Cz) and central-parietal (CPz) areas were the most salient.}
	\label{fig:Figure_3}
\end{figure}

\begin{figure}[htbp]
	\centering
		\includegraphics[width=\textwidth]{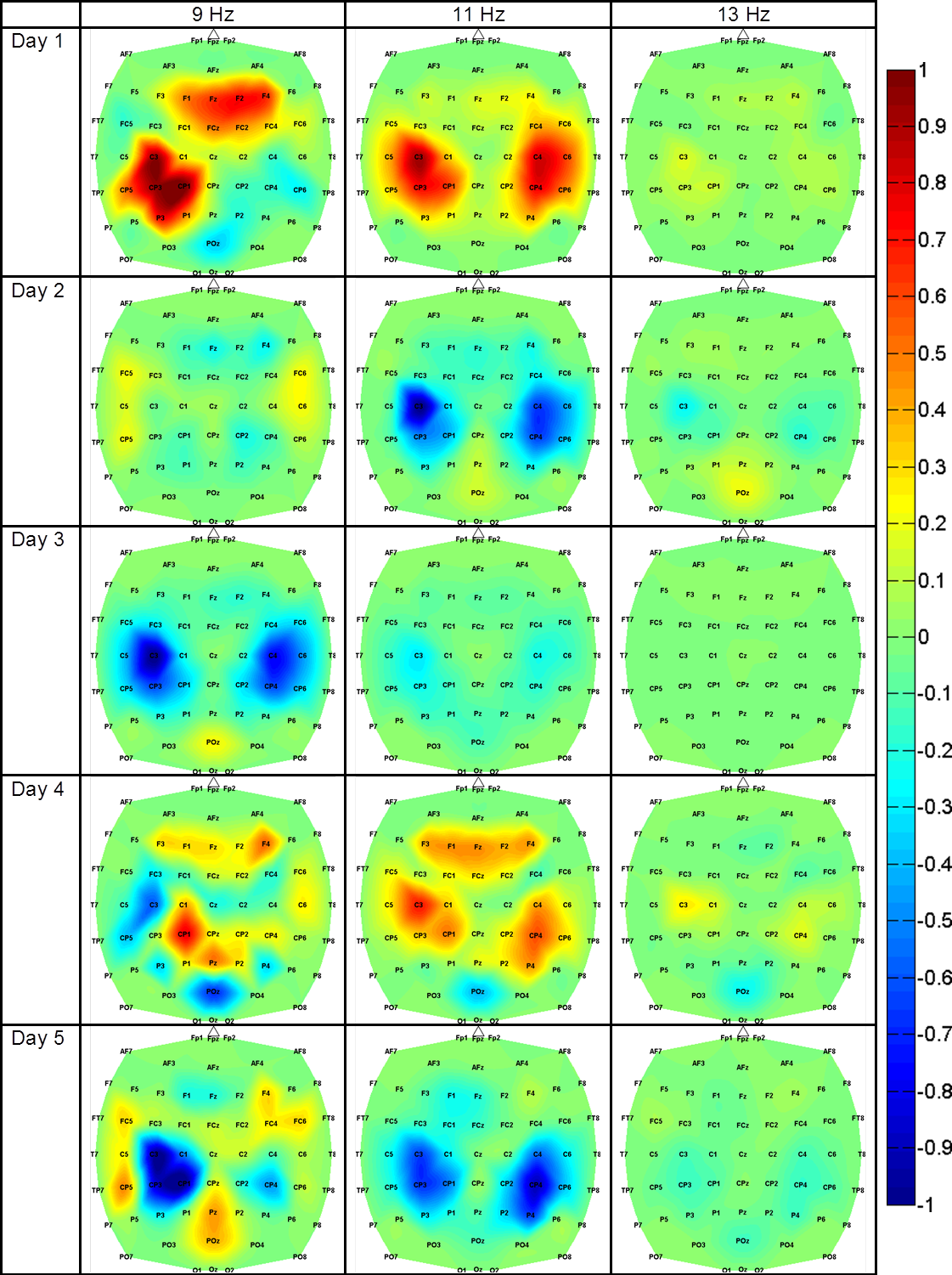}
	\caption{Feature extraction images of Subject 5 for all experimental sessions. The EEG power in the 9-13 Hz bins in the mid-frontal (Fz), lateral central (C3 and C4) and lateral central-parietal (CP3 and CP4) areas were the most informative for encoding the differences between idling and walking KMI.}
	\label{fig:Figure_4}
\end{figure}

\subsection*{Online Calibration}\label{sec:calibresults}
After the short calibration procedure (see \nameref{sec:oc} in the Methods section), the state transition thresholds, $T_{\mathcal{I}}$ and $T_{\mathcal{W}}$, were determined from the distributions of the posterior probabilities, $P(\mathcal{W}\,|f^{\star})$ (see Fig.~5 for an example), and their values are presented in Table 3. Note that in an ideal situation, $P(\mathcal{W}\,|f^{\star}\in\mathcal{W})$ and $P(\mathcal{W}\,|f^{\star}\in\mathcal{I})$ should cluster around 1 and 0, respectively, but as long as these probabilities are separable, online BCI control should be achievable. The values of $T_{\mathcal{I}}$ and $T_{\mathcal{W}}$ ranged from 0.07 to 0.70 and from 0.09 to 0.90, respectively, and the average across all subjects on all experimental days was 0.40 and 0.62 for $T_{\mathcal{I}}$ and $T_{\mathcal{W}}$, respectively.

\begin{figure}[htbp]
	\centering
		\includegraphics[width=\textwidth]{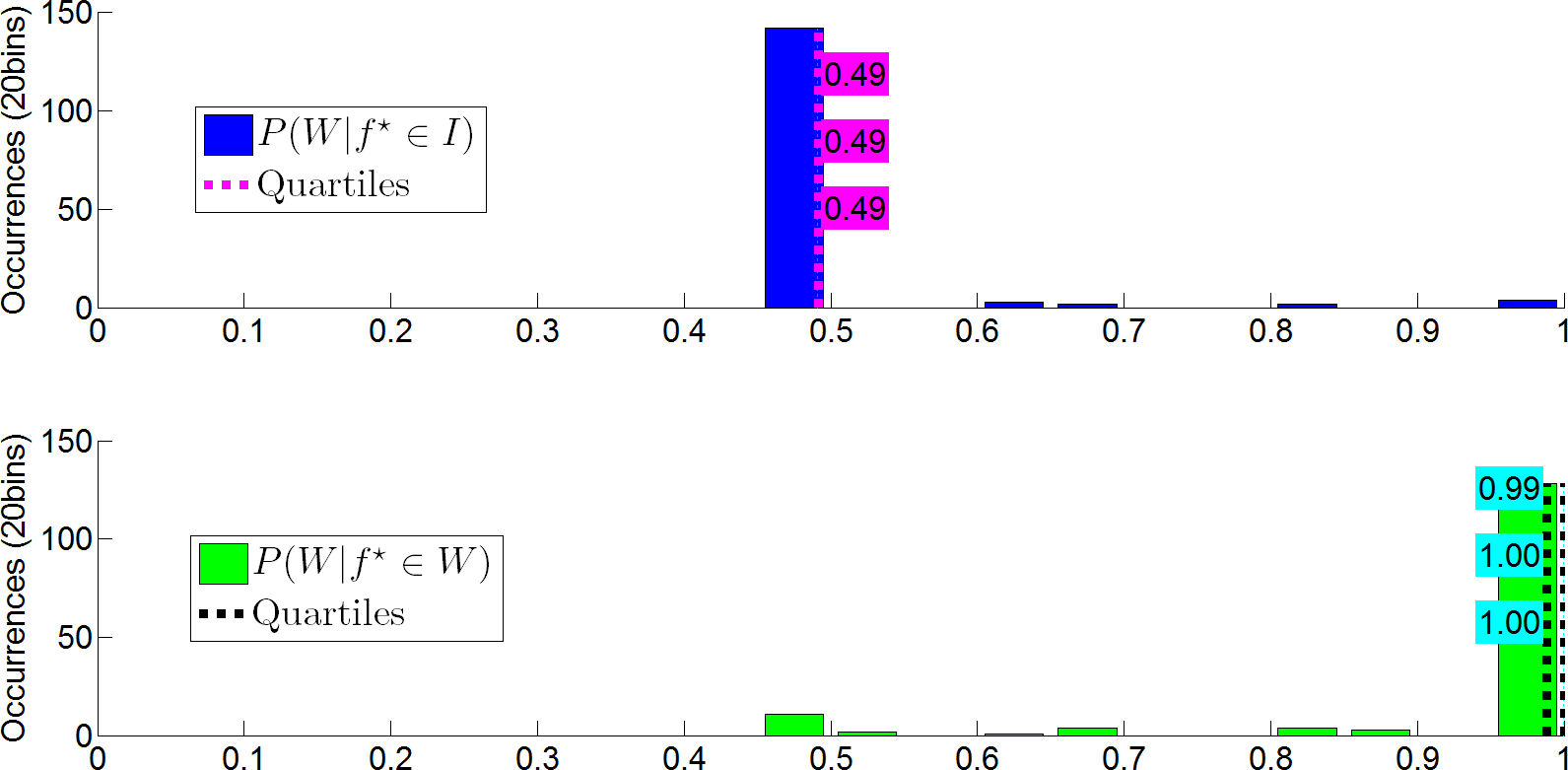}
	\caption{Histograms of the posterior probability of walking KMI given idling, $P(\mathcal{W}\,|f^{\star}\in \mathcal{I})$, and  walking KMI given walking, $P(\mathcal{W}\,|f^{\star}\in \mathcal{W})$, for Subject 3. Note that $P(\mathcal{W}\,|f^{\star}\in \mathcal{I})=1 - P(\mathcal{I}\,|f^{\star}\in \mathcal{I})$. Dashed lines indicate the 25\%, 50\%, and 75\% quantiles.}
	\label{fig:Figure_5}
\end{figure}

\begin{table}[htbp]
	\centering
\begin{tabular}{ccll}
\toprule
\multicolumn{4}{c}{Online Parameters} \\
\midrule
Subject  &  Day  &  $T_{\mathcal{I}}$  &  $T_{\mathcal{W}}$  \\
1    &  1  &  0.37  &  0.47  \\
     &  2  &  0.42  &  0.67  \\
     &  3  &  0.45  &  0.55  \\
     &  4  &  0.22  &  0.39  \\
     &  5  &  0.35  &  0.44  \\
\midrule
2    &  1  &  0.5   &  0.5   \\
     &  2  &  0.07  &  0.09  \\
     &  3  &  0.42  &  0.45  \\
     &  4  &  0.6   &  0.7   \\
     &  5  &  0.6   &  0.66  \\
\midrule
3    &  1  &  0.2   &  0.3   \\
     &  2  &  0.57  &  0.62  \\
     &  3  &  0.26  &  0.75  \\
     &  4  &  0.7   &  0.9   \\
     &  5  &  0.4   &  0.9   \\
\midrule
4    &  1  &  0.62  &  0.8   \\
     &  2  &  0.61  &  0.8   \\
     &  3  &  0.38  &  0.45  \\
     &  4  &  0.4   &  0.6   \\
     &  5  &  0.58  &  0.65  \\
\midrule
5    &  1  &  0.3   &  0.85  \\
     &  2  &  0.4   &  0.7   \\
     &  3  &  0.3   &  0.7   \\
     &  4  &  0.4   &  0.8   \\
     &  5  &  0.3   &  0.85  \\
\bottomrule
\end{tabular}
	\caption{The stop (and start) parameters, $T_{\mathcal{I}}$ (and $T_{\mathcal{W}}$), for online operation as determined by the calibration session.}
	\label{tab:table3}
\end{table}

\subsection*{Online Performances}\label{sec:onlineresults}
All subjects were able to achieve purposeful online control immediately (on day 1), with the exception of Subject 2 (on day 2). The online BCI performances upon achieving purposeful control are summarized in Table 4. The average stop score was 7.4 $\pm$ 2.3, and the average completion time was 273 $\pm$ 51 sec across all online sessions ($n=124$). Note that minimal lapses in BCI control occurred, as only 5 out of 124 online sessions ($\sim$4\%) were non-purposeful. Examples of the best online performances for each subject are shown in Fig.~6 along with the simulated random walk PDFs. For additional comparison, the average performance achieved by a manually controlled physical joystick was 9.38 $\pm$ 0.95, and on their best day, Subjects 3, 4, and 5, achieved a similar number of successful stops; however, no subjects were able to complete the course as fast as manual control. A representative time-space course of an online session (see Fig.~7) shows only a single false alarm  and a single omission. Also, a supplementary video showing a representative online session from Subject 3 is can be found at http://www.youtube.com/watch?v=K4Frq9pwAz8. 

\begin{table}[htbp]
	\centering
\begin{tabular}{llrrr}
\toprule
\multicolumn{2}{l}{Subject}   & Completion Time       & Successful Stops & Session    \\
       &                      & mean$\pm$std (sec)  & mean$\pm$std   & Breakdown  \\
\midrule
1      &    n=29              &        275$\pm$45   &  6.2$\pm$1.8   & (26, 3) \\ 
       &    Best: Day 5       &        298$\pm$77   &  6.8$\pm$2.3   &            \\
       &    Random Walk       &        258$\pm$12   &  6.9$\pm$1.3   &            \\
\midrule
2      &    n=25              &        271$\pm$66   &  5.7$\pm$2.3   & (24, 1) \\
       &    Best: Day 5       &        293$\pm$26   &  8.1$\pm$1.2   &            \\
       &    Random Walk       &       1050$\pm$85   & 10.0$\pm$0.2   &            \\
\midrule
3      &    n=24              &        277$\pm$65   &  9.4$\pm$1.3   & (24, 0) \\
       &    Best: Day 4       &        231$\pm$8    & 10.0$\pm$0.0   &            \\
       &    Random Walk       &              $>$1200  &  0.1$\pm$0.25  &            \\
\midrule
4      &    n=19              &        289$\pm$43   &  8.3$\pm$1.8   & (18, 1) \\
       &    Best: Day 1       &        264$\pm$12   &  8.9$\pm$0.3   &            \\
       &    Random Walk       &              $>$1200  &  4.3$\pm$0.7   &            \\
\midrule
5      &    n=27              &        258$\pm$31   &  7.7$\pm$2.1   & (27, 0) \\
       &    Best: Day 4       &        260$\pm$17   & 10.0$\pm$0.0   &            \\
       &    Random Walk       &              $>$1200  &  5.1$\pm$1.4   &            \\
\midrule
\multicolumn{2}{l}{All subjects}   & 273$\pm$51  & 7.4$\pm$2.3 & (119, 5) \\ 
\midrule
\multicolumn{2}{l}{Physical joystick} & 205.07$\pm$4.2 & 9.38$\pm$0.95 & \\
\bottomrule
\end{tabular}
	\caption{Average online performances, including completion time and number of successful stops for each subject, upon achieving purposeful control, which was on day 2 for Subject 2, and on day 1 for all other subjects. Also presented are the average online performances of each subject's best experimental day. Performances of random walk are shown for comparison. Sessions are also broken down by number of purposeful and non-purposeful performances, i.e. (p$<$0.01, p$\geq$0.01)}
	\label{tab:table4}
\end{table}

\begin{figure}[htbp]
	\centering
		\includegraphics[width=\textwidth]{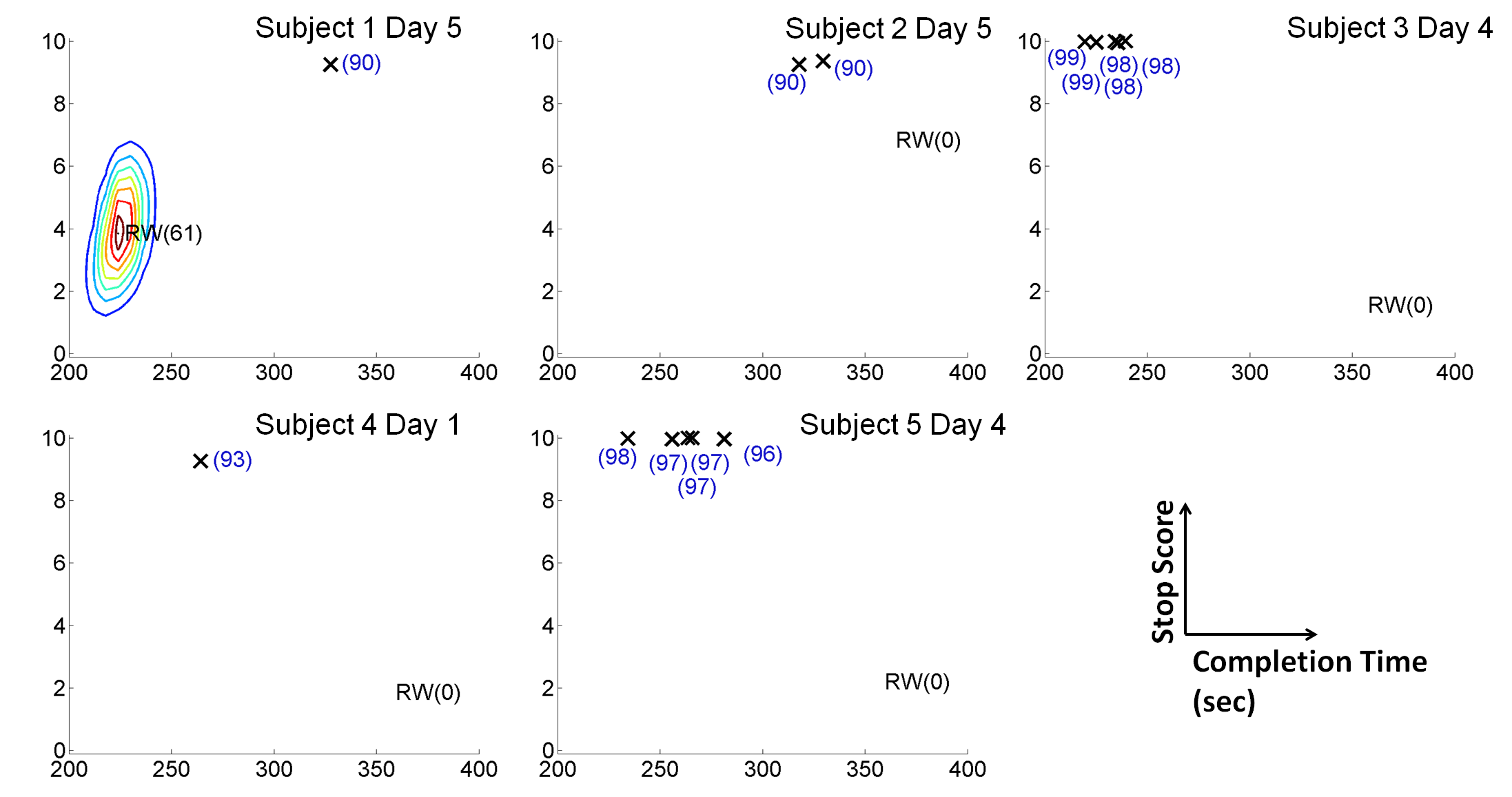}
	\caption{Best online performances (completion times and stop scores) of each subject. Performances are marked by crosses with associated composite scores shown in parenthesis. PDF of the random walk simulations (contour lines) are also shown for Subject 1, but are absent for other subjects as the contours lie outside of the alloted 20-min limit.  Note that all performances are purposeful (p$<$0.01).}
	\label{fig:Figure_6}
\end{figure}

\begin{figure}[htbp]
	\centering
		\includegraphics[width=1.00\textwidth]{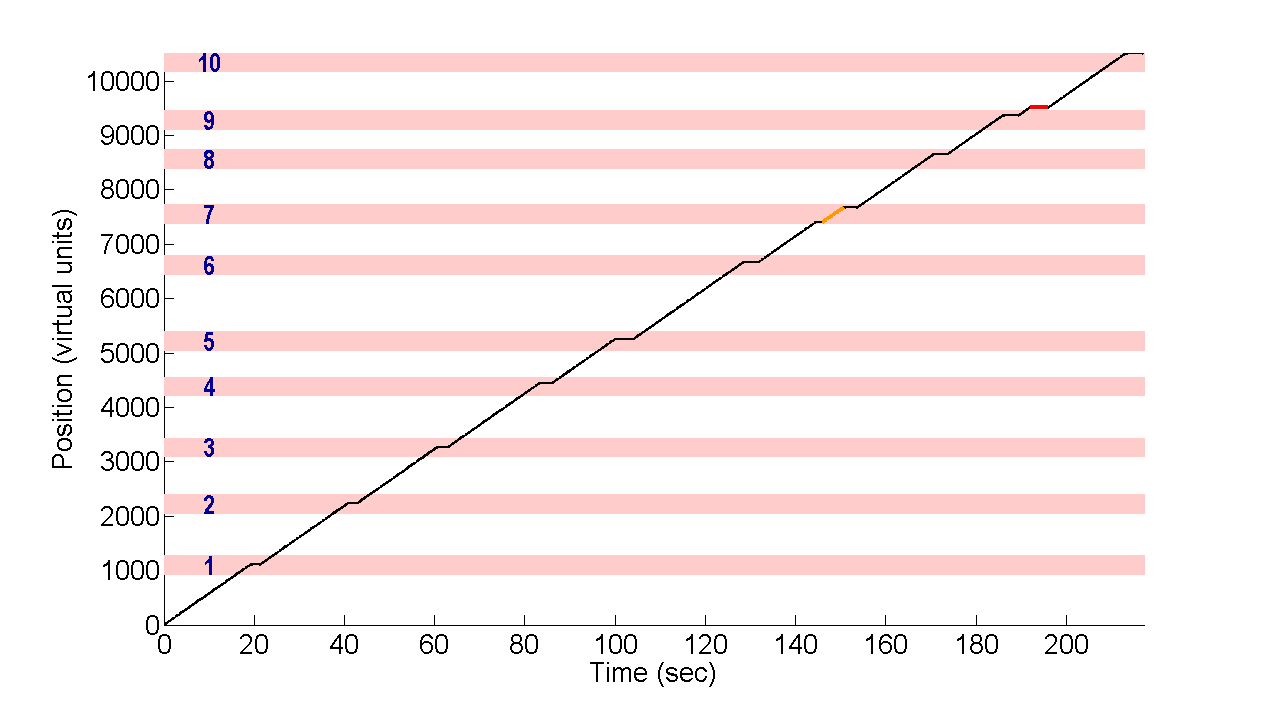}
	\caption{A representative time course of an online session where Subject 3 was able to achieve a high number of successful stops with a short completion time. False positives (i.e. the avatar walked when the subject intended to stop) are marked by orange segments, and false negatives (i.e. the avatar stopped when the subject intended to walk) are marked by red segments.}
	\label{fig:Figure_7}
\end{figure}

The composite scores were calculated from the online performance measures and are summarized in Table 5. In general, the performances improved significantly over time; the average on day 1 was 77.8\% and the average on day 5 was 85.7\% (p$=$0.0302). For comparison, the composite score of the joystick task is also shown. On their best days, Subjects 3, 4 and 5 achieved performances similar to those of the manually controlled joystick, reaching nearly perfect performances (100\%).    

\begin{table}[htbp]
	\centering
\begin{tabular}{cclllll}
\toprule
           &       & \multicolumn{5}{c}{Composite Score (\%)} \\
\cmidrule{3-7}
Subject    &       & Day 1           & Day 2           & Day 3           & Day 4           & Day 5           \\
\midrule
  1  &        &  66.2$\pm$3.1  &  76.8$\pm$9.1  &  75.0$\pm$11.3  &  74.4$\pm$11.6  &  76.9$\pm$11.9  \\
     &  Best  &          69.7  &             90  &           85.6  &           80.5  &          89.9  \\
\midrule
  2  &        &                &  69.9$\pm$11.2  &  69.3$\pm$13.5  &  59.9$\pm$12.3  &  85.4$\pm$5.4  \\
     &  Best  &                &           86.8  &           89.3  &           68.9  &          90.4  \\
\midrule
  3  &        &  86.4$\pm$3.2  &  87.5$\pm$14.2  &   93.0$\pm$5.6  &   98.4$\pm$0.4  &  97.3$\pm$2.2  \\
     &  Best  &          89.9  &           95.2  &           98.5  &             99  &          98.6  \\
\midrule
  4  &        &  91.5$\pm$1.7  &   90.1$\pm$5.3  &   89.7$\pm$4.1  &  79.4$\pm$16.3  &  80.7$\pm$2.3  \\
     &  Best  &          93.2  &           95.5  &           93.1  &           96.1  &          82.3  \\
\midrule
  5  &        &  66.4$\pm$9.0  &   80.7$\pm$5.1  &   93.5$\pm$3.5  &   96.9$\pm$0.9  &  88.7$\pm$6.1  \\
     &  Best  &          76.7  &           86.8  &           97.3  &           98.2  &          98.3  \\
\midrule
Average &     & 77.8$\pm$13.0  &  79.8$\pm$11.5  &  81.7$\pm$13.9  &  80.7$\pm$17.7  &  85.7$\pm$10.2 \\
\midrule
Joystick &    &                &                 &  96.5$\pm$3.8   &                 &                \\
\bottomrule
\end{tabular}
	\caption{Average and best composite online performance score for all subjects for each experimental day as calculated using Eqs.~\ref{eq:composite2}~and~\ref{eq:composite1}.  The composite scores for  random walk and the physical joystick are also shown for comparison. Note that Subject 2 did not achieve purposeful control on day 1 and was therefore unable to participate in online sessions on this day. }
	\label{tab:table5}
\end{table}

\section*{Discussion}
The results of this study show that subjects with paraplegia or tetraplegia due to SCI can operate a non-invasive BCI-controlled avatar within a VRE to accomplish a goal-oriented ambulation task. All subjects gained purposeful online BCI control on the first day after undergoing a 10-min training session, with the exception of Subject 2, who did not attain control until day 2. In addition, BCI control was maintained and continued to improve over the course of the study. These findings suggest that a BCI-controlled lower extremity prosthesis for either gait rehabilitation or restoration may be feasible. 

The offline classification accuracies varied across subjects and experimental days, but were significantly above the chance level performance (50\%). Similar to able-bodied subjects engaged in the same task~\cite{ptwang:10, ptwang:12}, a short 10-min training session was sufficient for the data-driven machine learning algorithm to generate accurate subject-specific EEG decoding models for this population. The topographic maps of these models (e.g. Fig.~3 and Fig.~4) showed that the spatio-spectral features underlying the differences between walking and idling KMIs varied across subjects and evolved over experimental days. The differences in the brain areas and EEG frequencies across subjects may be due to variations in cortical reorganization following SCI~\cite{Cramer2005, Sabbah2002, Alkadhi2005, Hotz-Boendermaker2008}, or due to differences in imageries employed by each subject (e.g. the KMI of walking instructions may have been interpreted differently by each subject). Nevertheless, all subjects showed activation of mid-frontal areas, which likely overlay the pre-motor and supplementary motor areas, as well as the pre-frontal cortex. Their activation during walking KMI is consistent with functional imaging findings, such as those in~\cite{clafougere:10}. Another common pattern across subjects was the presence of activity near bilateral, lateral central-parietal electrodes, which likely represents the arm sensorimotor areas. A similar pattern was observed in able-bodied individuals~\cite{ptwang:12}, and is hypothesized to originate from arm swing imagery. Finally, the evolution of the feature extraction maps over the 5 experimental days may be indicative of a neuro-plasticity process associated with practice and learning~\cite{nsward:03,sjpage:09}. 

The spatio-temporal variations of walking KMI activation patterns demonstrate the necessity of a data-driven machine learning approach for rapid acquisition of intuitive BCI control. First, our approach accommodates for the variations of these activity patterns across subjects, as well as their evolution over time. Second, it facilitates rapid acquisition of online BCI control, presumably by enabling subjects to utilize intuitive mental strategies. The user training time necessary to acquire purposeful BCI control in this study is significantly shorter than those of other BCI studies where users must learn a completely new cognitive skill to modulate pre-selected EEG features, such as the $\mu$-rhythm over lateral central areas~\cite{jrwolpaw:04}. Finally, this approach carries a significant potential value in the future practical implementation of BCI-prostheses, as it may drastically reduce the training time needed to attain purposeful and useful BCI control from a timescale of weeks to months to one of minutes to days. This in turn may significantly reduce the cost of training users to operate future BCI-prostheses.

The results presented in Table 4 show that once purposeful control was achieved, it was maintained in 96\% of all online sessions. In addition, 3 out of 5 subjects achieved successful stop scores similar to those obtained using a manually controlled joystick. Even though no subjects were able to complete the course as fast as manual control, it is encouraging that the average composite scores increased significantly over the course of the study. Furthermore, the average composite scores (Table 5) improved over time, with the best scores approaching 100\% for Subjects 3 and 5 by the end of the study. Therefore, not only was online control significantly different from random walk, but it was also meaningful. Given this trend, additional training and practice may help further improve performance, possibly to the point of approaching that of the manually controlled joystick. In conclusion, the high level of online control achieved by SCI subjects over the course of 5 experimental days suggests that it may be feasible to apply this BCI system to control a lower extremity prosthesis for ambulation after SCI. Furthermore, the proposed BCI-VRE system may serve as a training platform for operation of BCI lower extremity prostheses once they become widely available.

\section*{Conclusions}
This study shows that SCI subjects can purposefully operate a self-paced BCI-VRE system in real time, and that the current BCI design approach is able to overcome the potential problems associated with variations in neurophysiology due to cortical reorganization after SCI, learning and plasticity processes, and differences in KMI strategies. Furthermore, the system satisfies the requirements of an ideal BCI-lower extremity prosthesis set forth in~\cite{ptwang:12}, namely: intuitiveness, robustness, and short training time. The operation of the system is intuitive as it enabled subjects to use walking KMI 
to control the ambulation of the avatar. The system is robust in that the data-driven decoding methodology was able to successfully accommodate for subject-to-subject and day-to-day variations in the neurophysiological underpinnings of idling and walking KMI behaviors. In addition, subjects were able to maintain purposeful online control over the course of several weeks, further underscoring the system's robustness over time. Finally, the system required only a short training time, as BCI control was generally attained after only a 10-min training data collection procedure followed by a 2-min calibration session on the 1$^{\text{st}}$ experimental day (for 4 out of the 5 subjects). The successful outcome of this study indicates that BCI-controlled lower extremity prostheses for gait rehabilitation or restoration after SCI may be feasible in the future. 

\bigskip

\section*{List of abbreviations}
SCI, Spinal cord injury; BCI, Brain-computer interface; VRE, Virtual reality environment; KMI, Kinesthetic motor imagery; EEG, Electroencephalogram; FES, Functional electrical stimulation; EMG, Electromyogram; FFT, Fast Fourier Transform; CPCA, Classwise principal component analysis; LDA, Fisher's linear discriminant analysis; AIDA, Approximate information discriminant analysis; 1D, One-dimensional; CV, Cross-validation; NPC, Non-player character; MC, Monte Carlo; PDF, Probability density function; ASIA, American Spinal Injury Association. 

\section*{Competing interests}
CEK received salary from HRL Laboratories, LLC (Malibu, CA). The remaining authors declare that they have no competing interests.

\section*{Author's contributions}
 
CEK carried out the experiments, collected and analyzed the data, and wrote the article. PTW programmed the brain-computer interface software, assisted with carrying out the experiments, collecting the data, and analyzing the data. LAC contributed to conception of the study. AHD conceived and designed the study, implemented the VRE, recruited and consented subjects, supervised the experiments, and co-wrote and proofread the article. ZN conceived and designed the study, designed the signal processing, pattern recognition, and classification algorithms, and co-wrote and proofead the article. All authors read and approved the final manuscript.
    
\section*{Acknowledgments}
  \ifthenelse{\boolean{publ}}{\small}{}
        This study was funded by the Roman Reed Spinal Cord Injury Research Fund of California (RR 08-258 and RR 10-281), and was partially funded by the Long Beach VA Advanced Fellowship Research Award.
 
\newpage
{\ifthenelse{\boolean{publ}}{\footnotesize}{\small}
 \bibliographystyle{bmc_article}  
  \bibliography{SCIgait} }     


\ifthenelse{\boolean{publ}}{\end{multicols}}{}

\end{document}